# Accelerated hot-carrier cooling in MAPbI$_3$ perovskite by pressure-induced lattice compression

Loreta A. Muscarella[1], Eline M. Hutter[1,2], Jarvist M. Frost[3], Gianluca G. Grimaldi[1], Jan Versluis[1], Huib J. Bakker[1], Bruno Ehrler[1,*]

[1] Center for Nanophotonics, AMOLF, Science Park 104, 1098 XG Amsterdam, the Netherlands

[2] Department of Chemistry, Utrecht University, Princetonlaan 8, 3584 CB, Utrecht, the Netherlands

[3] Department of Physics, Imperial College London, South Kensington, London SW7 2AZ, United Kingdom

\* **Corresponding author**

b.ehrler@amolf.nl

**Abstract**

Hot-carrier cooling (HCC) in metal halide perovskites in the high-density regime is significantly slower compared to conventional semiconductors. This effect is commonly attributed to a hot-phonon bottleneck but the influence of the lattice properties on the HCC behaviour is poorly understood. Using pressure-dependent transient absorption spectroscopy (fs-TAS) we find that at an excitation density below Mott transition, pressure does not affect the HCC. On the contrary, above Mott transition, HCC in methylammonium lead iodide (MAPbI$_3$) is around two times as fast at 0.3 GPa compared to ambient pressure. Our electron-phonon coupling calculations reveal about two times stronger electron-phonon coupling for the inorganic cage mode at 0.3 GPa. However, our experiments reveal that pressure promotes faster HCC only above Mott transition. Altogether, these findings suggest a change in the nature of excited carriers in the high-density regime, providing insights on the electronic behavior of devices operating at such high charge-carrier density.

**TOC**

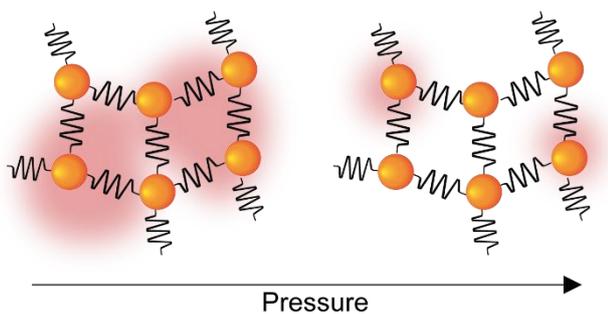

**Introduction**

Photoexcitation with a photon energy larger than the bandgap of semiconductors results in the formation of a non-thermal distribution of "hot charge-carriers" (*i.e.* high-energy electrons in the conduction band and high-energy holes in the valence band). In semiconductor solar cells, these charges relax to the conduction (and valence) band edge before they are collected. The first step is thermalization[1,2], occurring within a few hundred femtoseconds, where the generated hot-carriers interact with each other through carrier-carrier scattering until they reach a common quasi-Fermi temperature, usually much higher than the lattice temperature. Subsequently, HCC occurs though carrier-phonon or carrier-impurity scattering until a thermal equilibrium with the local lattice is reached, usually on picosecond timescales[1,3,4]. In this step, heat is dissipated in the lattice through acoustic phonons. In solar cells, 33% of the energy of sunlight is lost as heat[5,6,7,8,9] during the thermalization and cooling processes. Slow HCC is desired for thermoelectric devices[10], and hot carrier solar cells[11,12,13] where extracting carriers before they have cooled could enables breaking the thermodynamic limit for single-junction solar cells. Emissive applications such as lasers[14], single-photon sources[15,16] and optical modulators[17] require short HCC time for efficient radiative recombination and to prevent carrier trapping. In particular for lasers, it is quintessential to understand the electronic properties at the high-carrier density required to obtain lasing.

Metal halide perovskites have sparked great interest because of their unique properties such as a facile fabrication route, high defect tolerance, large absorption coefficient and remarkably low recombination rates, the latter similar to those of single-crystalline semiconductors. Also, the HCC timescale in this class of materials was found to be significantly slower[3,18–21] than in conventional semiconductors like InN[22] or GaAs[18] under the same (high) excitation density conditions. In $MAPbI_3$ perovskite, as in other semiconductors such as GaAs, the HCC time has been found to slow down upon increasing the charge-carrier density above a certain threshold[3,19,23,24]. This slower cooling has been explained from an effect that is commonly known as hot-phonon bottleneck. The origin of this phenomenon is still under debate, but it has been attributed to several mechanisms such as the accumulation of optical phonons that cannot be easily dissipated[18,25,3], optical-acoustic phonon up-conversion[20] and polaron formation[26–29].

Metal halide perovskites are polar semiconductors, and thus their electronic properties are expected to be strongly coupled with the lattice vibrations. Applying external pressure directly affects the lattice dynamics, and therefore can be used to tune properties that are strongly dependent on the lattice vibrations, such as the electron-phonon coupling and the phonon lifetimes. Changes in one or both of these quantities can affect the HCC. Increasing the electron-phonon coupling is expected to lead to faster HCC. Replacing iodide with a lighter halide increases the frequency of the longitudinal optical phonon mode ($\omega_{LO}$)[30] and thus shortens the HCC time measured under the same excess of energy and excitation density conditions[31,32,3]. In $MAPbI_3$, the acoustic phonon lifetime, responsible for heat transport, has been found in the range of a few ps[33,34,35], two orders of magnitude shorter than in conventional semiconductors[36], resulting in an acoustic phonon mean free path of around 10 nm as reported by Wang *et al.*[37]. Thus, the thermal transport at room temperature in $MAPbI_3$ is highly inefficient, leading to slower HCC as the optical phonon population will build up, and to strong local heating if no thermal management strategy is applied. If the acoustic phonon lifetime increases, the HCC is expected to become faster. Understanding how lattice properties relate with HCC provide insights on the electronic behaviour of devices operating at different charge-carrier density. Thus, an effective strategy to manipulate *ad hoc* the HCC time is required to design devices with a specific application (*e.g.*, longer cooling time is desired for thermoelectric devices but undesirable for single-photon sources) operating in a certain charge-carrier density regime.

In this work, we combine pressure-dependent femtosecond transient absorption spectroscopy (fs-TAS) and electron-phonon coupling calculations to elucidate the effect of lattice compression on the factors that influence the HCC. We use pressure-dependent fs-TAS to experimentally probe the HCC time in $MAPbI_3$ at room temperature at pressures ranging from 0 to 0.3 GPa at varying light intensities.

At low excitation density ($7 \times 10^{17}$ photons/cm$^3$), the HCC time is fast (0.3 - 0.5 ps) and independent of pressure. High excitation density (> $10^{18}$ photons/cm$^3$) triggers a Mott transition where, microscopically, the material undergoes a transition from an insulating exciton gas – where polarons act independently – to a metal-like electron-hole plasma, and where the thermal energy is shared between overlapping polarons. At an excitation density above the Mott transition ($5.9 \times 10^{18}$ photons/cm$^3$), the HCC is significantly slower (2-3 ps) at ambient pressure, but accelerates by a factor 2-3 upon increasing the pressure to 0.3 GPa. In solar cells and optoelectronic devices operating above the Mott transition (> $10^{18}$ photons/cm$^3$), faster HCC timescale may allow for a faster dissipation of heat and therefore a lower operating temperature.

**Result and Discussions**

Solution-processed MAPbI$_3$ thin films were deposited by spin coating onto quartz substrates as reported in the **Method Section.** Absorbance measurements as a function of pressure and X-Ray diffraction measurements were performed on the sample to confirm the bandgap energy and the high crystallinity of the sample (**Figure S1**). Pressure-dependent transient absorption measurements were performed inside a hydrostatic pressure cell filled with the inert hydraulic liquid tetradecafluorohexane (FC-72, see **Method Section**) as depicted in **Figure 1a**. A 100-fs pulsed pump beam, with an energy of 3.1 eV and represented in purple in the schematic, is used for photoexciting the sample whereas a 100-fs pulsed probe beam (white light), depicted in the schematic as combination of all the wavelengths of visible light, is used to probe the excitation-induced change in transmission of MAPbI$_3$ on a picosecond timescale. The two beams are overlapped on the sample inside the pressure cell, and the arrival time of the two pulses is controlled with a delay stage.

Photoexcitation with a photon energy larger than the MAPbI$_3$ bandgap lead to a population of high-energy carriers (electrons and holes) (**Figure 1b**, dark red curve) that undergo rapid (~ 85 fs[1]) thermalization, faster than our temporal resolution. This thermalization process results from carrier-carrier scattering. The resulting hot-carrier population can be described as a quasi-Fermi distribution. Once the hot-carrier population reaches a common temperature $T_c$ (**Figure 1b**, red curve), higher than the lattice temperature $T_L$, carrier-phonon interactions dominate the HCC until $T_c$ is in equilibrium with $T_L$ (**Figure 1b**, yellow curve).

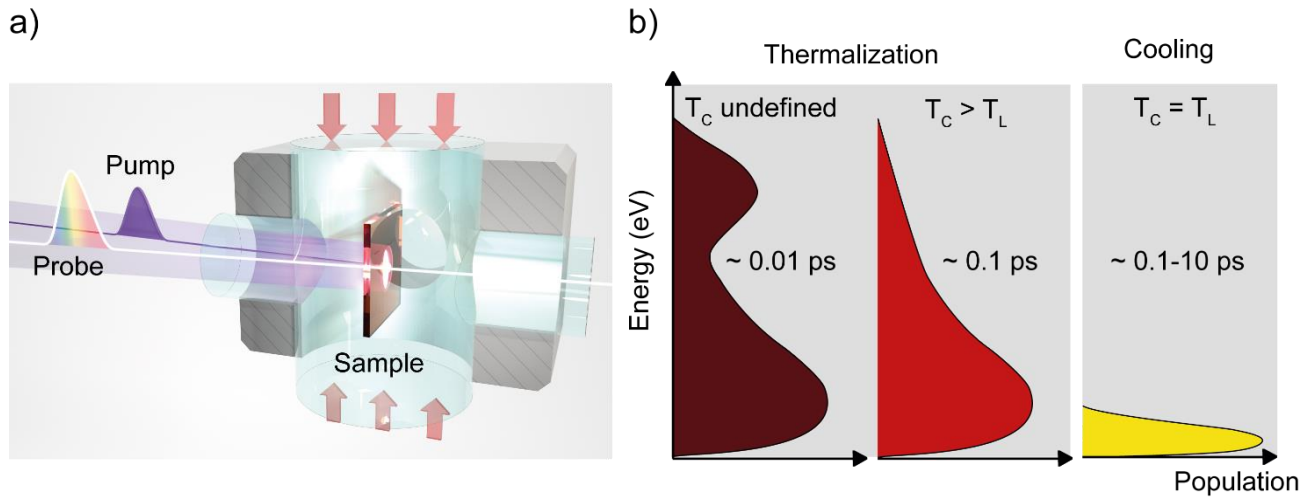

**Figure 1.** Schematic representation of the **a)** pressure-dependent fs-TA setup showing the generation of hot-carriers (white spot on the sample) and **b)** HCC mechanism following photoexcitation.

**Figure 2a-d** show representative ultrafast transient absorption traces for a pump–probe delay between 0-5 ps of MAPbI$_3$ photoexcited at 3.1 eV (bandgap 1.7 eV) with an initial carrier density $n_0$ of $7.0 \times 10^{17}$ photons/cm$^3$ (**Figure 2a,b**) and $5.9 \times 10^{18}$ photons/cm$^3$ (**Figure 2c,d**) at ambient pressure and at 0.3 GPa, respectively. The carrier density is calculated from the pump fluence, the fraction of absorbed photons at each pressure and the thickness of the perovskite film as measured by SEM cross-section, described in **Supplementary Note 1**. No degradation is observed at high excitation density as demonstrated in **Figure S2** from the stability of the TA signal over the course of the measurement. The $\Delta T/T$ traces show three features: (i) a positive $\Delta T/T$ centred at the bandgap energy of ~1.67 eV corresponding to the ground state bleach (GSB) signal that results from the band filling effect; (ii) a negative $\Delta T/T$ feature at energies below the bandgap (< 1.67 eV) at early times resulting from the bandgap decrease induced by the high energy carriers.[2]; (iii) a negative and broad $\Delta T/T$ signal at energies above the bandgap (> ~1.7 eV) resulting from light absorption of the photo-generated carriers. Pseudo-color TA plots of the same sample as a function of the pump-probe delay and probe energy are reported in **Figure S3.**

The hot-carrier population is evident from the width of the initial GSB signal, that shrinks over the course of the measurement (picosecond timescale) as the hot carriers cool down to the lattice temperature. This feature represents an average of the hot electron and hot hole temperatures given the effective masses are very similar[38].

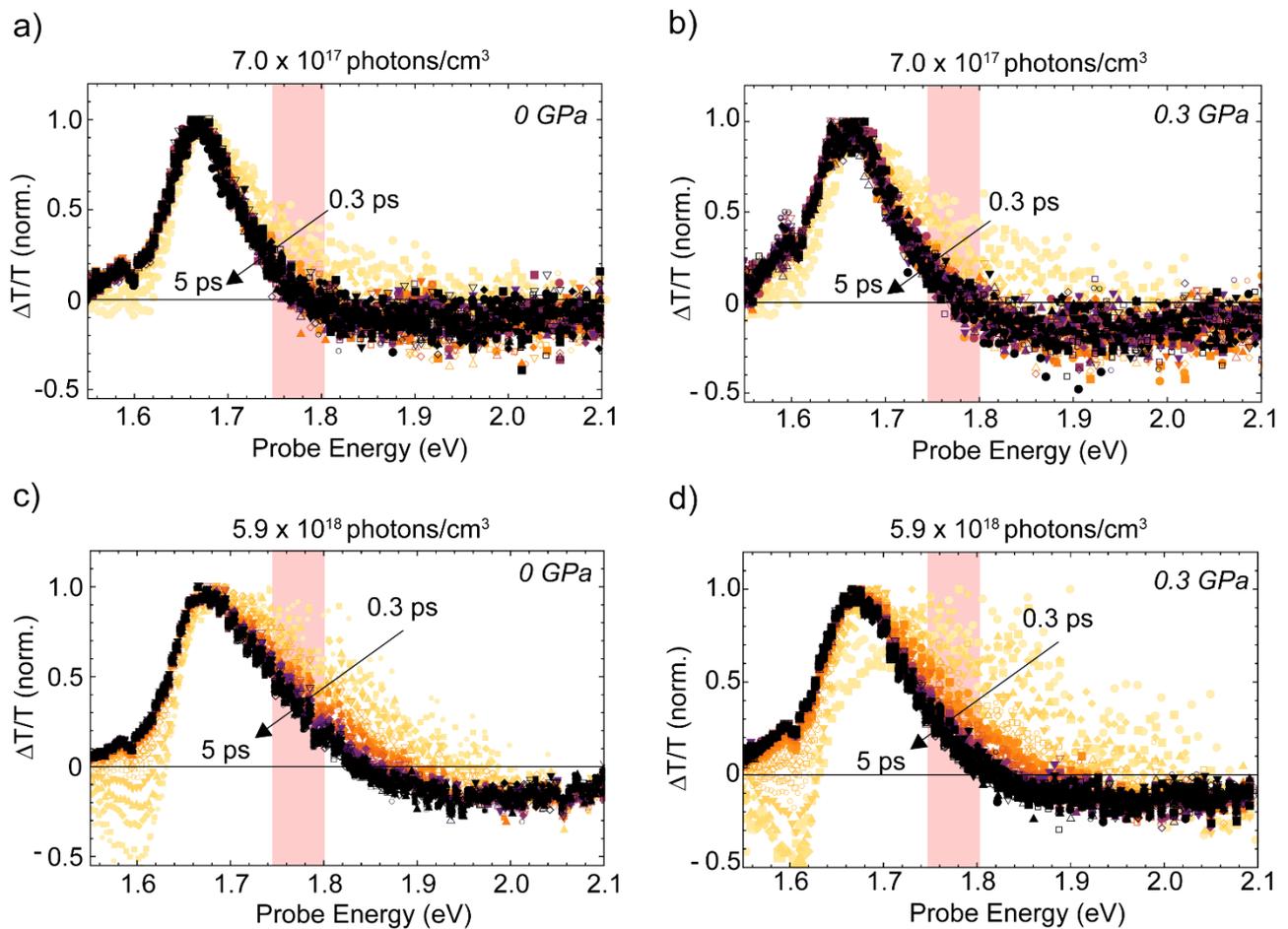

**Figure 2.** Charge carrier cooling as a function of hydrostatic pressure measured by pressure-dependent transient absorption spectroscopy. **a)** $\Delta T/T$ traces of MAPbI$_3$ with an initial carrier concentration of $7.0 \times 10^{17}$ photons/cm$^3$ at ambient pressure and **b)** at 0.3 GPa as a function of the probe energy. **c)** *ΔT/T* traces of MAPbI$_3$ with an initial carrier concentration of $5.9 \times 10^{18}$ photons/cm$^3$ at ambient pressure and **d)** at 0.3 GPa as a function of the probe energy.

A comprehensive model to obtain the HCC time from the transient absorption measurements has still to be developed but several methods are commonly used to obtain the HCC time and temperature[19,23,39]. We use two fitting strategies to obtain the trend of the HCC time as a function of pressure and excitation density. Both strategies yield a comparable HCC time upon increasing the pressure and for both excitation densities. The first method consists of integrating the Δ*T*/*T* in the region of the high-energy GSB tail from 1.75 – 1.80 eV and plotting the result as a function of delay time. The integrated range is highlighted in red in **Figure 2**. We plot the integrated traces for the high-density regime in **Figure 3a**. We fit the traces in the low- and high-density regime with a convolution of the instrumental response function (IRF) and an exponential decay function (see **Supplementary Note 2** for the analytical function and **Figure S4** for the IRF fit). At low excitation density ($7 \times 10^{17}$ photons/cm$^3$) no pressure dependency is observed, showing fast (~ 0.3 ps) HCC at all pressures investigated (**Figure 3b,** blue). At high excitation density, the decay comprises a fast component with a time constant of a few picoseconds (attributed to the HCC) and a slow component with a time constant in the order of tens of picoseconds. The presence of an additional slower process in the high-density regime has been shown previously and its origin is still under debate[16,19,20,40], therefore we compare the two fast components at the two excitation density used. The time constant for the short-lived component in the high-density regime is plotted in **Figure 3b** (red). Contrary to the experiment at low excitation density, the experiment conducted on the same sample at high excitation density shows almost three times faster HCC (time constant ~1 ps) at 0.3 GPa compared to ambient pressure (~3 ps).

To make sure the extracted trend of the HCC time with pressure is not affected by the energy range integrated, we performed the same fit but integrating Δ*T*/*T* in various energy ranges of the broad tail (**Figure S5** and **Figure S6** for the high- and low-density regime, respectively). The absolute values of the HCC time constants slightly vary, but the trend as a function of pressure is consistent. The second method used to obtain the HCC time consists of approximating the high-energy tail of the GSB and the negative PIA[20] with a modified Maxwell–Boltzmann distribution function[19,41,42], $f_{MB}(x) = \frac{A_1}{x} \sqrt{x - E_g}\, Exp\left(-\frac{x-E_f}{kT_c}\right) + \frac{A_2}{\sqrt{x-E_g}}$. Here, $A_1$ and $A_2$ represent the amplitudes of the band-filling and bandgap renormalization components, $k$ is the Boltzmann constant, $E_g$ is the bandgap energy as calculated from the fit of the GSB as a function of pressure at later times when the HCC is completed (**Figure S7**), $E_f$ is the electron (and holes) quasi-Fermi level, and $x$ is the energy. We approximate $E_f$ as $E_g$ and obtain the values of $A_1$, $A_2$ and $T_c$ from the fit. The second term of this fitting function is included to fit the negative feature originating from hot carrier induced bandgap reduction. We perform this fitting method on the same data. To probe the same population of hot carriers at all pressures conditions, the selected starting energy of the fit is the one corresponding to 1/2 of the maximum bleach. To ensure to probe the hot-carrier population after thermalization (but before HCC is finished), we only consider the data starting from 0.5 ps after photoexcitation. The fit yields the carrier temperatures which we show in **Figure 3c** as a function of delay time for ambient and two representative high-pressure conditions. The initial temperature depends on the excess energy of the photoinduced carriers, and the excitation density. This is the temperature reached by the carriers directly after the thermalization. Interestingly, we observe a higher initial hot-carrier temperature upon increasing the pressure. The slight redshift in the bandgap energy (7 meV, **Figure S7**) when we increase the pressure cannot account for this effect. We thus conclude that pressure may have an effect on the thermalization process as well, but because this process is faster than our temporal resolution we cannot further investigate this effect. The cooling time plotted in **Figure 3d** for the low- and high-density regime (in blue and red, respectively) is obtained by fitting an exponential decay function to the curve in **Figure 3c** taking into account the temperature errors of each datapoint. The HCC at ambient pressure is much slower than at high pressure, as reflected in a longer HCC time. The absolute values obtained by this fitting procedure for the cooling time are slightly lower than those obtained with the integration method shown in **Figure 3b**. As before, in the low-density regime, HCC is fast, and there is no variation within the experimental error. In the high-density regime we find again that the HCC is two to three times faster between 0 and 0.3 GPa, the same result obtained with the integration model.

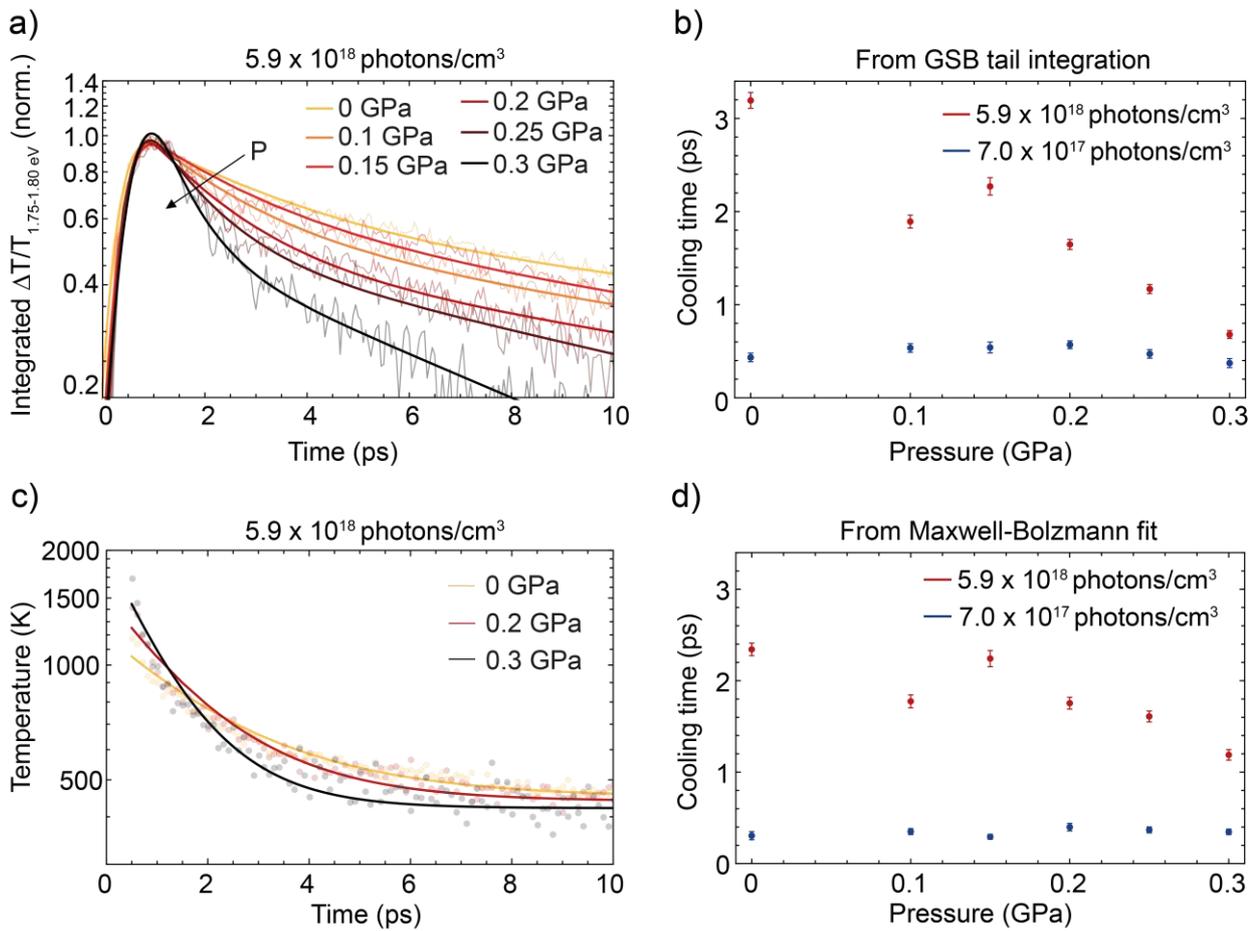

**Figure 3.** Two fitting strategies for determining the HCC time as a function of pressure. **a)** Normalized $\Delta T/T$ integrated in the range of 1.75-1.80 eV in the GSB tail as a function of time, and delay time for a high excitation density of 5.9 x $10^{18}$ photons/cm$^3$ at different applied pressures. The decay of the GSB tail becomes faster upon increasing pressure. **b)** HCC time extracted from the fit of the tail decay as a function of pressure at a high excitation density (red) and a low excitation density

regime (blue). **c)** Carrier temperature obtained by fitting the high-energy tail of the GSB with a Maxwell-Boltzmann distribution, as a function of delay time for a high excitation density of 5.9 x $10^{18}$ photons/cm$^3$ at different applied pressures, and **d)** HCC time obtained from the fit of the temperature decay as a function of pressure in the high- (red) and low-density regime (blue).

It is well-established that the HCC time depends on experimental parameters like the ambient temperature, the excess of excitation energy compared to the bandgap of the material, and the excitation density[3,27,43]. We can exclude significative changes in these parameters (see **Supplementary Note 3** for details) upon changing the pressure, and therefore we can attribute the pressure-dependent trend observed solely to changes in the material properties following compression, in particular in the electron-phonon coupling.

In order to understand what is happening on the microscopic scale within the material as a function of pressure, we calculate the electron-phonon coupling for all the phonon modes. The results of the relevant phonon modes are shown in **Figure 4a**. In these calculations we uniformly increase the pressure in a semi-local DFT electronic structure calculation (see **Method section**), starting with an ambient pressure pseudo-cubic MAPbI$_3$ structure[44]. The number of the phonon modes shown in **Figure 4a** is the index in ascending energy order. The modes 1-3 represent the acoustic modes, that play a negligible role in the initial HCC when charge carriers strongly couple to optical modes though Fröhlich interactions. Modes 4-6 and 7-9 with frequencies ranging from 19 to 35 cm$^{-1}$ (0.5-1 THz) represent octahedral twist and distortion, respectively. These phonon modes generally preserve the bond length and have limited coupling with the organic cation, in contrast to the phonon modes with frequencies above 65 cm$^{-1}$ (~ 2 THz), from mode 10 onwards. For these latter modes the change in bond length leads to collisions with the A$^+$-site ion, inducing coupling with the tumbling MA$^+$ [45]. Modes 10-16 are coupled motions of the organic cation with the inorganic sub-lattice and rotational vibration of the cation around the nitrogen or the carbon atom. Modes 17-36 are related to intramolecular vibration of the organic cation, and not relevant in the context of HCC as the electronic modes are localized mostly on the inorganic cage. A detailed assignment and description of the phonon modes is reported in Leguy et al.[46] Previous works[47,48] have shown that the dominant phonon mode coupled to the excited state dynamics observed by fs-TAS in MAPbI$_3$ has a frequency around 27 cm$^{-1}$ (~ 0.8 THz), and that there is a somewhat less strongly coupled mode at higher frequencies. For this reason, we confine our discussion to phonon modes 4-9 as these are the most relevant for the coupling with the electrons at the conduction band minimum. Whereas most of phonon modes show no clear trend with pressure, the electron-phonon coupling of mode 5 (~27 cm$^{-1}$, ~ 0.8 THz) associated with the octahedral twist, highlighted in grey in **Figure 4a**, shows a significant and approximately linear increase (**Figure 4b**) when pressure rises from ambient to 1 GPa, with two times enhancement at 0.3 GPa compared to ambient pressure.

A quantitative prediction of HCC (a phenomenological quantity) from the calculation of microscopic electron-phonon coupling matrix elements requires a mechanistic model of the cooling processes, and a detailed consideration of the electron-phonon interaction across the double electron and phonon Brillouin zones. This is beyond the scope of the present paper.

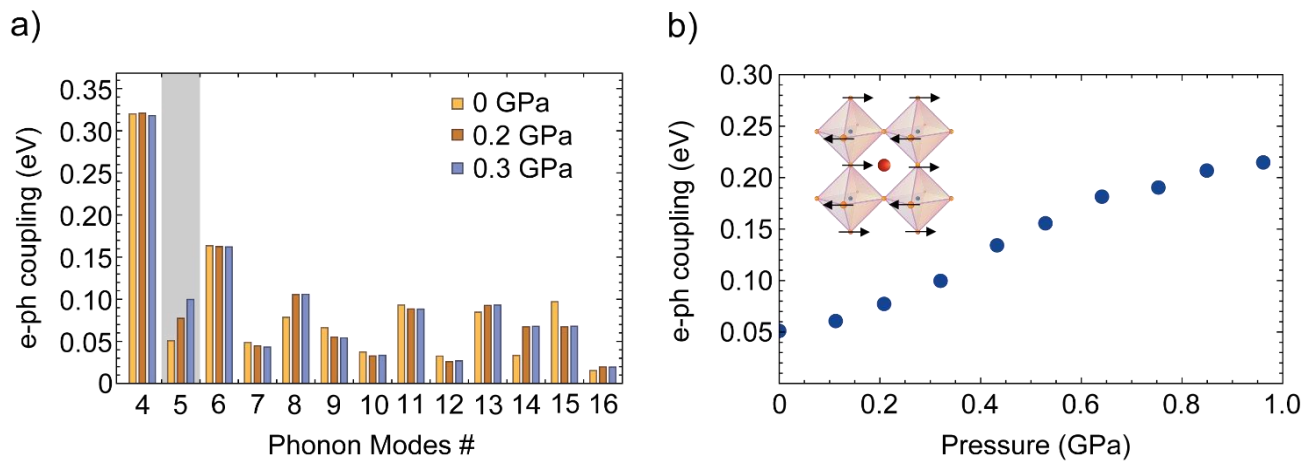

**Figure 4. a)** Electron-phonon coupling strength of the MAPbI$_3$ phonon modes to the conduction band minimum (CBM) at the gamma point in the Brillouin zone in a 2x2x2 supercell, listed in ascending phonon energy. Highlighted in grey the phonon mode 5 (~27 cm$^{-1}$, ~0.8 THz), affected most strongly

by pressure. **b)** Electron-phonon coupling of the mode 5 (~27 cm$^{-1}$, ~0.8 THz) as a function of pressure. *Inset*: atomic motions related to the phonon mode 5 (octahedral twist).

One would anticipate that the greater electron-phonon coupling at higher pressure would result in faster HCC, both in the low- and the high-density regime. However, we only observe a dependence on pressure at high densities. In the low-density regime, below the Mott transition, HCC is fast (0.3-0.5 ps) and independent of pressure. In a polar material, the dielectric electron-phonon coupling dominates because it is long-range[49], and will provide the main channel by which electrons photo-excited above-bandgap will lose their energy. This coupling involves the interaction of the charge of the carriers with the transition dipole moment of the surrounding phonon modes. The variational polaron method predicts a polaron relaxation time constant of ~ 0.1 ps[50] for this material, which is consistent with our observation of a very short HCC time. This long-range dielectric coupling is not expected to be significantly affected by pressure[51], in agreement with our observations.

Above the threshold for the Mott transition, previously calculated to be around $3 \times 10^{18}$ photons/cm$^3$ [28], the polarons overlap. In this high-density regime we propose that this dielectric coupling is screened, and instead the HCC proceeds via the weaker local electron-phonon coupling. We calculate that this local electron-phonon coupling is proportional to pressure over the pressure range 0-0.3 GPa (in fact we find a linear trend up until 1.0 GPa), and so the HCC at high-density becomes faster with pressure. This result is related to what has been found by Mohanan et al.[52] who attributed the presence of a hot-phonon bottleneck to the deposition of a large portion of the initial energy on the optical modes at 27 cm$^{-1}$ (~0.8 THz). These optical phonon modes do not efficiently dissipate the excess energy, as they are isolated from the rest of the lattice and thus do not strongly interact with the rest of the phonon bath. Our findings thus reveal that pressure enhancement of the electron-phonon coupling can be used to manipulate HCC at densities above the Mott transition density, while having no effect in the low-density regime where the long-range dielectric electron-phonon coupling dominates.

The hot-phonon bottleneck in MAPbI$_3$ perovskite has also been attributed to the extremely short acoustic phonon lifetime[53] of this material, which also causes its low thermal conductivity[54]. After the interactions between electrons and optical phonons, the latter decay into acoustic phonons that transport heat through the lattice. A short acoustic phonon lifetime can thus be responsible for the suppression of heat dissipation, as this energy could be reabsorbed by optical phonons and thus create a hot-phonon bottleneck that slows down the cooling of the hot electron-phonon plasma. Although this mechanism occurs in the second stage of the cooling (tens of picosecond), it might have some minor influence on the first cooling stage as well. To determine to what extent the faster HCC results only from the enhanced electron-phonon coupling or from a combination with a longer acoustic phonon lifetime would require an expensive computation that is beyond the scope of this work.

In conclusion, we used pressure-dependent fs-TAS to investigate the effect of external pressure on hot-carrier cooling in MAPbI$_3$ thin films. We found that in the low-density regime (below the Mott transition) the HCC time is not affected by pressure, whereas it becomes two to three times faster in the high-density regime (above the Mott transition). Our calculations reveal a twofold enhancement of the electron-phonon coupling for the mode related to the octahedra twist when the pressure is increased from ambient pressure to 0.3 GPa. These findings, together with the observed difference in the behaviour in the low and high-density regime, suggest the presence of two different mechanisms dominating HCC at the two excitation densities explored. In the low-density regime, below the Mott transition where polarons do not overlap, the long-range dielectric electron-phonon coupling dominates. In the high-density regime, above the Mott transition, this contribution is suppressed as the polarisation fields of the polarons overlap forming an electron-hole plasma and the HCC occurs via local electron-phonon coupling. This local contribution is significantly weaker at ambient pressure leading to slow HCC but increases linearly over the 0.3 GPa pressure range studied. These findings contribute to the understanding of how applied stress can be used to control the HCC time in halide-perovskite devices for emissive applications such as lasers and single-photon sources.

**Supporting Information**

Absorption and XRD spectra of $MAPbI_3$ thin films; $\Delta T/T$ as a function of the probe energy and real time for $MAPbI_3$ at ambient pressure for degradation; 2D plot of $\Delta T/T$ as a function of the probe energy and time for $MAPbI_3$ at ambient pressure and 0.3 GPa; IRF as a function of pressure calculated from $MAPbI_3$ excited resonantly at 800 nm; cooling time extracted from the integration of $\Delta T/T$ in various energy ranges of the broad tail as a function of pressure in the low- and high-density regime; $MAPbI_3$ bandgap as a function of pressure; excitation density calculation; fitting function for the $\Delta T/T$ decay in the high-energy tail; influence of experimental conditions in the HCC time;

**Author contributions**

L.A.M. performed the pressure-dependent TA experiments and data analysis under the supervision of B.E. J.V. assisted in the TA experiments under the supervision of H.J.B. G.G.G and E.M.H gave input on the data analysis and the design of the experiment. J.M.F. performed the electronic-structure calculations. The manuscript was written by L.A.M and J.M.F. with input from all authors.

**Experimental Methods**

**Sample fabrication**. $MAPbI_3$ precursor solution (1.05 M) is prepared by mixing MAI (TCI, >99%) and lead iodide (TCI, 99.99%, trace metals basis) in N,N-dimethylformamide (DMF, Sigma Aldrich anhydrous, ≥99%) in a 1:1 ratio. Thin films are prepared by spin coating the precursor solution at 9000rpm for 30s and the antisolvent (Chlorobenzene, Sigma Aldrich, anhydrous, ≥99%) dripped after 15s on quartz. Films are subsequentially annealed at 100 °C for 1 hour. The precursor solution preparation and spin coating are conducted in a nitrogen-filled glovebox.

**Characterization.** The XRD pattern of perovskite films deposited on quartz was measured using an X-ray diffractometer, Bruker D2 Phaser, with Cu K$\alpha$ ($\lambda$ = 1.541 Å) as X-ray source, 0.01° (2$\theta$) as the step size, and 0.100 s as the exposure time. Absorption spectra of MAPbI$_3$ films on quartz as a function of pressure were measured in a pressure cell (ISS Inc.) with a LAMBDA 750 UV/Vis/NIR Spectrophotometer (Perkin Elmer) from 550 nm to 850 nm, whereas absorbance spectra at ambient pressure are measured in ambient atmosphere in an integrating sphere.

**Pressure-dependent transient absorption spectroscopy.** Hydrostatic pressure was generated inside a pressure cell (ISS Inc.) increasing the volume of an inert liquid (FC-72, 3M) using a manual pump. Prior using, the liquid was degassed in a Schlenk line to remove oxygen which causes, from 0.3 GPa onwards, scattering of a fraction of light and therefore a reduction of the transmitted signal from the sample. The pressure was applied from 0.3 GPa to ambient pressure in steps of 0.050 GPa. The error in the pressure reading is estimated to be 0.020 GPa. We wait 7 minutes in between pressures for equilibration of the material under pressure. The transient absorption (TA) setup used for pressure-dependent measurements has been previously described in Hutter et al.[55] In addition, a reflective neutral density filter (OD 1) is placed in the probe path, before the sample, to reduce the 800 nm residue. To avoid polarization effects, the relative polarization of the probe and pump pulses was set to the magic angle (54.6°). The pump beam (3.1 eV) and the probe beam (white light) are overlapped inside the pressure cell during the measurement and the probe spot size was chosen to be smaller than the pump spot size to obtain homogenous excitation over the probed area. A mechanical delay stage is used to change the pump-probe overlap time to follow the evolution of the hot-carrier cooling from 0.05 ps to 30 ps with a step size of 0.05 ps. To correct all spectra for the chirp of the white light, we extract the chirp from the cross-correlation measured on a bare quartz substrate, inside the pressure cell and in the same conditions as the experiments with MAPbI$_3$.

**Semi-local density functional theory electronic structure calculation.** The reference structures at pressures 0 GPa to 1 GPa were generated with a Limited-memory Broyden–Fletcher–Goldfarb–Shanno (LBFGS) optimisation, using Perdew–Burke–Ernzerhof (PBE) exchange correlation functional with a plane-wave basis of 700 eV cut-off and a 6x6x6 Brillouin-zone integration grid. Pressure was applied in increments of 0.1 GPa. Electron-phonon couplings were then calculated by a linear combination of atomic orbitals (LCAO) method with a double-zeta basis set, in a 2x2x2 supercell of the pressurised realisations, with the PBE functional. All calculations were in Grade Point Average Weighted (GPAW)[56,57] using the Atomic Simulation Environment (ASE)[58] interface.

**Acknowledgments**

The work of L.A.M., E.M.H., J.V., H.J.B. and B.E. is part of the Dutch Research Council (NWO) and was performed at the research institute AMOLF. The work of L.A.M. was supported by NWO Vidi grant 016.Vidi.179.005. The authors thank Henk-Jan Boluijt for the design of Figure 1a. The authors thank María C. Gélvez-Rueda for commenting on the manuscript. J.M.F. is supported by a Royal Society University Research Fellowship (URF-R1-191292). Electronic structure calculations used the Imperial College Research Computing Service (DOI: 10.14469/hpc/2232).